\documentclass{article}
\usepackage{cite,graphicx,subfigure,setspace}
\usepackage[small,bf,up]{caption}

\begin{document}
\title{Observation of Transparency of Erbium-doped Silicon nitride in photonic crystal nanobeam cavities}
\author{Yiyang Gong$^{1}$, Maria Makarova$^{1}$, Sel\c{c}uk Yerci$^{2}$, Rui Li$^{2}$, \\Martin J. Stevens$^{3}$, Burm Baek$^{3}$, Sae Woo Nam$^{3}$, Luca Dal Negro$^{2,4}$ and Jelena Vu\v{c}kovi\'{c}$^{1}$ \\
    \small\textit{$^{1}$Department of Electrical Engineering, Stanford University, Stanford, CA 94305} \\
    \small\textit{$^{2}$Department of Electrical and Computer Engineering, Boston University, Boston, MA 02215, USA} \\
    \small\textit{$^{3}$National Institute of Standards and Technology, 325 Broadway, Boulder, CO 80305, USA} \\
    \small\textit{$^{4}$Division of Material Science, Boston University, Boston, MA 02215, USA}}

\maketitle
\begin{abstract}
One-dimensional nanobeam photonic crystal cavities are fabricated in an Er-doped amorphous silicon nitride layer. Photoluminescence from the cavities around 1.54 $\mu$m is studied at cryogenic and room temperatures at different optical pump powers. The resonators demonstrate Purcell enhanced absorption and emission rates, also confirmed by  time-resolved measurements. Resonances exhibit linewidth narrowing with pump power, signifying absorption bleaching and the onset of stimulated emission in the material at both 5.5 K and room temperature. We estimate from the cavity linewidths that Er has been pumped to transparency at the cavity resonance wavelength. 
\end{abstract}


The interest in combining electronics and optics has sparked a large effort to fabricate light emitting devices with silicon complementary metal-oxide-semiconductor (Si-CMOS) compatible materials. One possible material system for this application is Er-doped amorphous silicon nitride (Er:SiN$_{x}$), which emits at the telecom wavelength of 1.54 $\mu$m \cite{Yerci_SNerb,Rli_SNerb}. The Er emission can be sensitized by the host through a nanosecond-fast energy transfer mechanism from the amorphous nitride matrix (SiN$_{x}$), which provides an absorption cross-section four orders of magnitude larger than that of Er in silica (SiO$_{2}$) \cite{Yerci_SNerb,Rli_SNerb}. Low field electrical injection in this material is also possible, as demonstrated by electroluminescence of silicon nanocrystals in silicon-silicon nitride superlattices \cite{Negro_el,Yerci_el}.

In order to explore the possibility to achieve stimulated emission in this system, we couple emission from Er to photonic crystal (PC) cavities with high quality ($Q$-) factor and low mode volume ($V_{mode}$), as the interaction between the emitter and cavity mode can be tailored with design. The Purcell effect ($\propto Q/V_{mode}$) \cite{Purcell} in such cavities leads to enhanced spontaneous emission rates into the cavity mode and thus decreases the lasing threshold. We have already demonstrated enhancement of Er photoluminescence (PL) in Er doped silicon nitride coupled to two-dimensional (2D) silicon PC cavities, including linewidth narrowing of the PC cavity mode and Purcell enhancement of the Er emission rate \cite{Maria_ErbCav,Erbcav_gain}. Althought silicon-based PC cavities have high $Q$ and small $V_{m}$, the overlap of the cavity mode with the active material (Er-doped nitride cladding on silicon cavities) is small. In addition, absorptive losses, stemming mostly from the Si portion of the membrane, also limit the gain. Here we report on a PC cavity design made entirely of the Er:SiN$_{x}$ material with improved mode overlap with the active material and reduced absorptive losses. We observe two times larger linewidth narrowing relative to silicon PC cavities with Er-doped nitride, indicating a larger gain coefficient in the cavity \cite{Erbcav_gain}. Moreover, some cavities were pumped to transparency.

While the high indices of refraction of Si and GaAs ($n > 3$) have enabled high-$Q$ 2D PC cavities, there have been numerous recent efforts to  develop high-$Q$ PC cavities in low-index materials such as diamond ($n=2.4$) \cite{Hu_diamond,Kreuzer_diamond}, silicon nitride ($n=2.0$) \cite{Painter_1Dmodes,Marko_SiN1D}, and silicon dioxide ($n=1.5$) \cite{Gong_quartz1D}. Two-dimensional PC cavities confine light by distributed Bragg reflection (DBR) in the 2D PC plane and total internal reflection (TIR) in the surface normal direction. However, because it is difficult to achieve a large 2D photonic band in low index-contrast material systems \cite{Maria_SiNCPC}, it is preferable to design low index photonic crystal cavities in a one-dimensional geometry, relying on DBR in the direction along a narrow beam, and total internal reflection in the other two directions. In particular, the cavity is formed by modifying the size and spacing of several holes at the center of the beam, thus forming linear or parabolic optical potential wells. Quality factors as high as $10^5$ have been experimentally achieved in silicon nitride by employing such designs \cite{Painter_1Dmodes}.

\begin{figure}[htbp]
\centering
\includegraphics[width=5.5in]{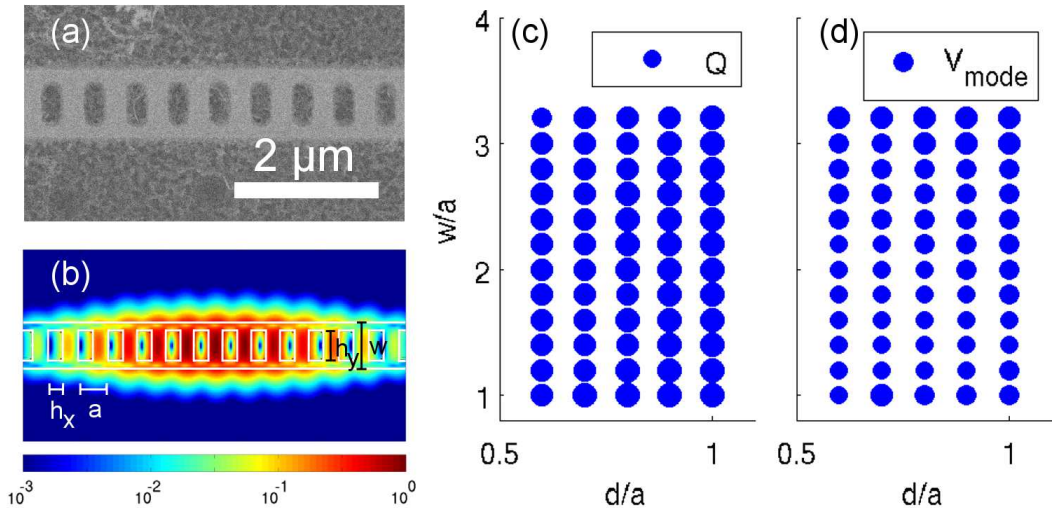}
\caption{(a) Scanning electron micrograph (SEM) of the fabricated Er:SiN$_{x}$ nanobeam cavity. (b) The $|E|^2$ profile of the fundamental cavity mode from FDTD simulations. The area of each marker illustrates the (c) $Q$ and (d) $V_{mode}$  of the cavity as the width and height of the beam is changed, while $h_{x}=0.5a$, $h_{y}=0.7w$, and the design of the holes are fixed. The reference markers represent $Q=30,000$ and $V_{mode}=0.95(\lambda/n)^3$.}
\label{fig:beam}
\end{figure}

In this work, we apply the parabolic design \cite{Painter_1Dmodes,Gong_quartz1D} to the Er:SiN$_{x}$ material, which has an index of refraction approximately the same as that of SiN$_{x}$ ($n=2.05$). The hole spacing at the center of the cavity is 0.88$a$, where $a$ is the lattice constant of the PC mirror outside of the cavity. The beam has thickness $d=0.8a$ and width $w=1.5a$. The width of the rectangular holes in the direction along the beam is $h_{x}=0.5a$, and the width perpendicular to the beam is $h_{y}=0.7w$ (Figure \ref{fig:beam}(b)). We employ 3D finite-difference time-domain (FDTD) simulations to calculate the field profile of the fundamental transverse-electric (TE)-like mode, as shown in Fig. \ref{fig:beam}(b). The mode has theoretical normalized frequency $a/\lambda=0.36$, quality factor $Q=30,000$, with mode volume $V_{m}= 0.95 (\lambda/n)^3$. In addition, the mode overlap $\Gamma$, defined as the fraction of the electric field energy in the active material, is $\Gamma = 52$\%, in the structures that have the active material distributed throughout the beam (i.e., the whole beam composed of Er:SiN$_{x}$),  which is 12 times improved relative to a hybrid Er:SiN$_{x}$/Si membrane \cite{Erbcav_gain}.

We also vary the beam width ($w$) between 1.0$a$ and 3.2$a$ and the beam thickness ($d$) between 0.6$a$ and 1.0$a$, fixing $h_{x}=0.5a$, $h_{y}=0.7w$, and the same design of holes for the cavity, and find $Q$ and $V_{mode}$ for the cavities. We observe that the $Q$ of the cavity has little dependence on the width of the beam, but does increase with the beam thickness (Fig. \ref{fig:beam}(c)). This is the case as SiN$_{x}$ has a fairly high index of refraction, and beam widths in the studied range can still support waveguide modes. In addition, we find that $V_{mode}$ is minimized around $w/a=1.6$ for various beam thicknesses (Fig. \ref{fig:beam}(d)). As expected, there is a tradeoff between the $Q$-factor and the mode volume of the resonator.

\begin{figure}[htbp]
\centering
\includegraphics[width=5.5in]{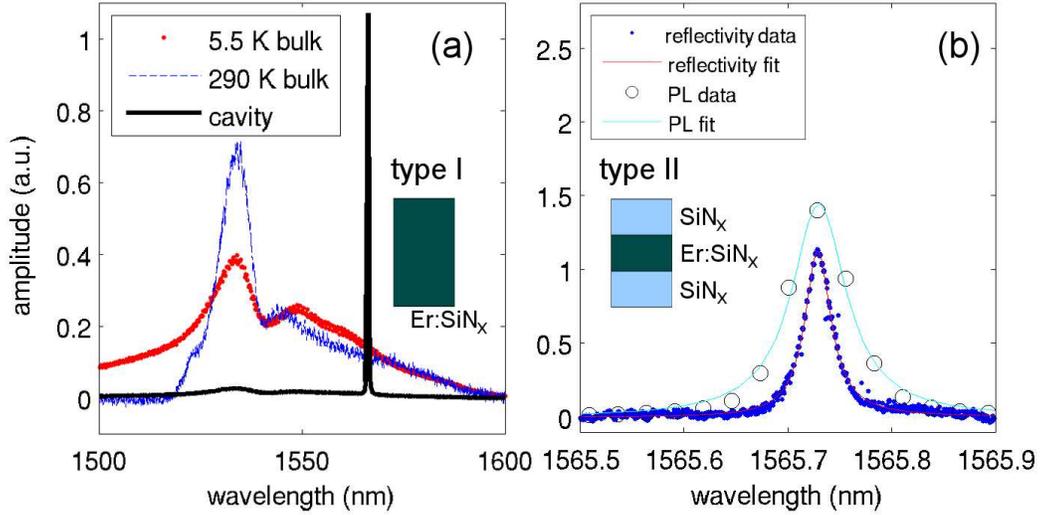}
\caption{(a) Photoluminescence from the cavity at room temperature and the unpatterned film at room temperature and 5.5 K. The whole membrane is composed of Er:SiN$_{x}$ in this case (type I, shown in inset). (b) Spectrum of a cavity fabricated in SiN$_{x}$ with only the middle third doped with Er (type II, shown in inset). Dots correspond to the spectrum obtained by a laser scan in cross-polarization reflectivity, and circles to PL measured by the spectrometer. Fits to a Lorentzian lineshape gives a $Q=52,000$ from the reflectivity scan and a spectrometer resolution limited $Q=25,000$.}
\label{fig:spectrum}
\end{figure}

Er:SiN$_{x}$ is grown on top of an oxidized silicon wafer by N$_{2}$ reactive magnetron co-sputtering from Si and Er targets in a Denton Discovery 18 confocal-target sputtering system, as discussed elsewhere \cite{Yerci_SNerb,Rli_SNerb}. The sacrificial oxide layer is 700 nm thick, while the Er:SiN$_{x}$ layer is 500 nm thick with an Er concentration of $3.0 \times 10^{20}$ cm$^{-3}$ (referred to as type I in the inset of Fig. \ref{fig:spectrum}(a)). Type II samples, shown in the inset of Fig. \ref{fig:spectrum}(b), have only the middle third of the membrane doped with Er. The growth is followed by a post-annealing process in a rapid thermal annealing furnace at 1180 $^{\circ}$C for 480 s under forming gas (5\% H$_{2}$, 95\% N$_{2}$) atmosphere. The fabrication of the resonators employs electron beam lithography with 400 nm of ZEP-520A as the resist. The written pattern is then etched into the Er:SiN$_{x}$ or SiN$_{x}$ slab with a CHF$_{3}$:O$_{2}$ chemistry. Finally, suspended PC membranes can be formed by undercutting the oxide layer with a 6:1 buffered oxide etch, and further undercutting of 3 $\mu$m of the silicon substrate with a XeF$_{2}$ etcher. The scanning electron micrograph (SEM) of the fabricated beam with a width of $w=1.5a$ is shown in Figure \ref{fig:beam}(a).

As in previous work, we pump the Er $I_{15/2}\rightarrow I_{11/2}$ transition at 980 nm in order to reduce the total material losses \cite{JSTQE_silicon,Erbcav_gain}. Micro-photoluminescence ($\mu$-PL) is performed from normal incidence for both the pump and the collection beams with a 100$\times$ objective lens with numerical aperture $NA=0.5$, and the emission is directed to an InGaAs CCD spectrometer. The bulk PL is shown in Fig. \ref{fig:spectrum}(a) for both room and 5.5 K temperatures. In the cavity PL we observe the two TE-like modes, and we choose to work with the first order (fundamental) mode, as it has the lowest mode volume. A sample cavity PL spectrum is shown in Fig. \ref{fig:spectrum}(a), and typical cavities have $Q>12,000$. 

In addition, we also fabricate larger cavities with the same design parameters, except with $w=2.5a$ and only the middle one-third of the slab doped with Er (type II in the inset of Fig. \ref{fig:spectrum}(b)). In such a cavity, we measure a $Q$ of 25,000 from PL taken with the spectrometer (Fig. \ref{fig:spectrum}(b)). However, since the corresponding linewidth of 0.06 nm is limited by spectrometer resolution, a scan of the cavity is performed in cross-polarization reflectivity \cite{Hatice_ccavs,Dirk_refl} in steps of 0.002 nm with a tunable laser. The reflectivity scan data are also shown in Fig. \ref{fig:spectrum}(b), and yields a $Q$ of 52,000.

\begin{figure}[htbp]
\centering
\includegraphics[width=3in]{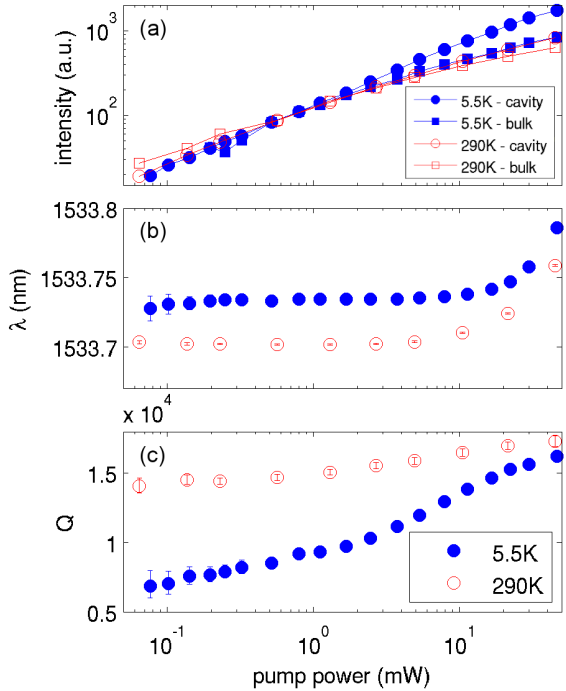}
\caption{The pump power dependence of the (a) integrated PC cavity intensity and PL spectrally decoupled from the cavity, (b) the cavity resonance wavelength, and (c) the cavity $Q$, all at 5.5K and 290K. The shift in wavelength between the two temperatures is most likely due to a shift of the sample position in the cryostat as temperature is varied. The pump power is measured in front of the objective lens.}
\label{fig:power}
\end{figure}

We perform power series measurements on one cavity (type I) at both room and cryogenic temperatures, with pump powers varying from 0.030 mW to 40 mW. We plot the integrated intensity from the cavity, and the integrated intensity of the uncoupled PL from the main Er emission lobe at 1525 nm $-$ 1540 nm (excluding the cavity, as the cavity and main Er emission lobe overlap), at both room temperature and 5.5K (Fig. \ref{fig:power}(a)). The amplitudes of the cavity and spectrally decoupled PL both increase sublinearly at both temperatures. However, it is interesting to note that the PL from the spectrally decoupled regions have approximately the same saturation characteristics at both temperatures, while the PL from the cavity resonance rise with a higher slope on the log-log plot than that of their uncoupled counterparts at both temperatures. Such behavior suggests a faster spontaneous emission rate at the cavity resonance, due to Purcell enhancement of radiative emission rate. In addition, at both temperatures, the cavities redshift at high pump power, suggesting cavity heating (Fig. \ref{fig:power}(b)). Finally, Fig. \ref{fig:power}(c) shows that at 5.5 K, the $Q$ increases dramatically with pump power from 6,000 to over 15,000. On the other hand, at room temperature, the $Q$ only increases from 14,000 to 16,000. In our previous work with an Er:SiN$_{x}$ layer coupled to a Si 2D PC cavity, we observed a smaller increase in $Q$ at cryogenic temperatures and negligible increase in $Q$ at room temperature \cite{Erbcav_gain}. As previously, here we attribute the difference in the behavior of $Q$ at the two temperatures to the broadening of the homogeneous linewidth of the Er emission transition with increasing temperature, which degrades Purcell enhancements of emission and absorption \cite{Erbcav_gain}. The larger changes in $Q$ in the nanobeam cavity (relatively to the change in \cite{Erbcav_gain}) at both temperatures are the result of greater mode overlap with the active material and the reduction of free carrier absorption, as the 980 nm pump is absorbed more by Si than by SiN$_{x}$. Because of the redshift in the cavity resonance wavelength with increasing pump power, we conclude that heating mechanisms due to free carriers or Er-Er interactions may still lead to increased absorption at high pump powers. Nevertheless, the reduced cavity losses of the new cavity design have enabled the observation of linewidth narrowing at room temperature.

\begin{figure}[htbp]
\centering
\includegraphics[width=3in]{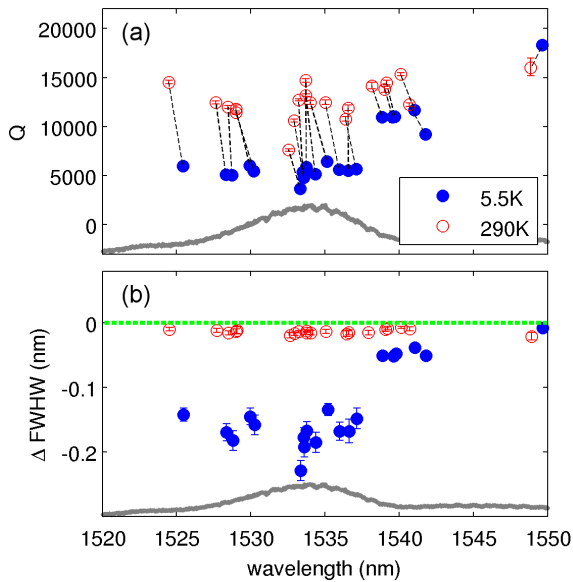}
\caption{(a) The $Q$s of the cavities at 5.5K and 290K, both with low pump power (less than 10 $\mu W$). The dashed lines connect the data for the same cavity at the two different temperatures. The shift in wavelength between the two temperatures is most likely due to a shift of the sample position in the cryostat as temperature is varied. (b) The change in the linewidth (full-width at half-max, FWHM) for individual cavities as pump power is switched from less than 10 $\mu W$ to 40 mW, at 5.5 K and room temperature. The scaled and shifted Er spectrum is shown in gray as a reference.}
\label{fig:cavities}
\end{figure}

By changing the PC lattice constant, and while maintaining the same cavity design, we fabricate cavities with a variety of wavelengths that span the Er emission spectrum. At room temperature, we observe that all cavities have $Q$s at or above 10,000, at high or low pump power (Fig. \ref{fig:cavities}(a)). However at low temperature, all the cavities that overlap with the main Er emission lobe demonstrate lower $Q$s at low pump power (below 10 $\mu$W pump). Then, by increasing the pump power, we observe a decrease in cavity linewidth and an increase in $Q$ (Fig \ref{fig:cavities}(b)), much like in our previous work \cite{Erbcav_gain}. However, the linewidth narrowing observed in this work is more than double that of our prior work, as expected, because of the higher mode overlap and reduced free carrier absorption. The maximum observed linewidth narrowing is by 0.23 nm at 5.5 K. In addition, the change in linewidth decreases with increasing cavity wavelength to almost negligible amounts. The spectral dependence of the linewidth narrowing is expected, as the wavelength range from 1525 nm to 1540 nm corresponds to the peak of Er absorption, and the absorption decreases with increasing wavelength. As opposed to our prior work \cite{Erbcav_gain}, here we observe linewidth narrowing for all cavities at room temperature as well (Fig \ref{fig:cavities}(b)). Although the linewidth changes at room temperature are much smaller than the changes for the same cavities at 5.5 K, they are still significant. At room temperature, the linewidth change is nearly the same for all cavity wavelengths. This can be explained by the large homogeneous linewidth of the Er at room temperature (up to 10 nm in glass hosts), such that a significant portion of the Er population couples to the cavity resonance. We have previously found that the $Q$s of the cavities, and thus the Er homogeneous linewidth, vary smoothly between 5.5 K and room temperature \cite{Erbcav_gain}. In addition, as the system is in the bad emitter limit (as we discussed previously in Ref. \cite{Erbcav_gain}), the absorption and emission rate enhancement depend only on the mode volume $-$ not the cavity Q factor. Thus, we have explored the two extremes of the temperature (5.5 K and 290 K) in order to display the largest variation in linewidth (at low pump power).

\begin{figure}[htbp]
\centering
\includegraphics[width=5.5in]{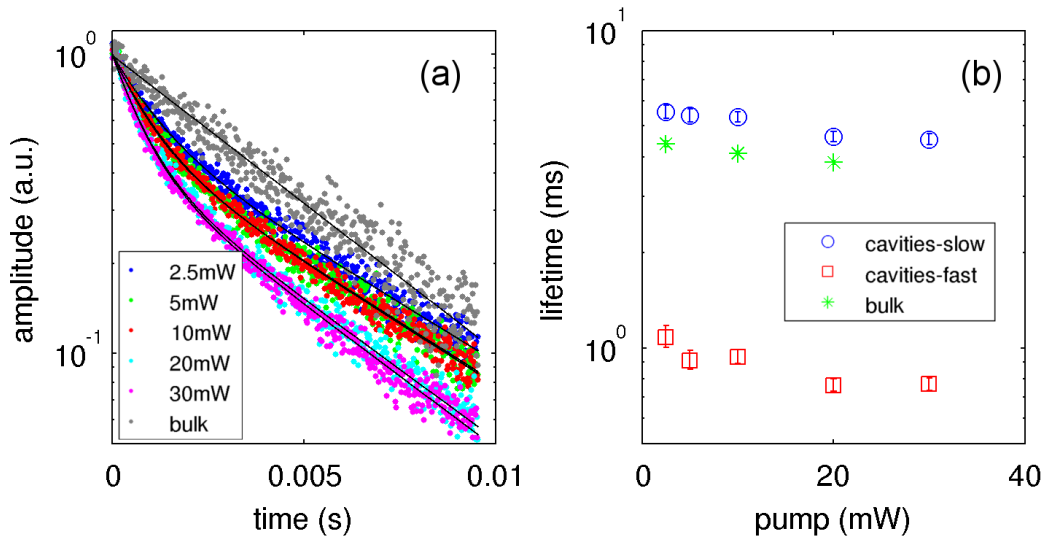}
\caption{(a) Time-resolved PL measurements of the cavity resonance for various pump powers at $\sim$3 K, as well as unpatterned film (integrated for all wavelengths). Solid lines for the cavity time traces are fits to a bi-exponential model. (b) The fast and slow components from the fits in part (a), as well as for an unpatterned film lifetimes for various pump powers.}
\label{fig:lifetime}
\end{figure}

Time-resolved PL measurements are performed using the same setup as in our previous work \cite{Erbcav_gain}. In summary, a 980 nm pump laser is chopped at 50 Hz by a mechanical chopper, and the decay of the PL is sent to a superconducting nanowire single-photon detector (SNSPD), which is held at a temperature of $\sim$3 K in a closed-cycle helium cryocooler \cite{SNSPD1}. The chopper provides start pulses to the time-sampling electronics, while the SNSPD provides a stop pulse each time it detects a photon. The electronics record a histogram of the number of stop counts arriving in  each 20 $\mu$s time interval after a start pulse; this histogram is proportional to the time response of the PL to the square wave pump. Time-resolved PL for the cavity design with $w=2.5a$ and the type II Er:SiN$_{x}$ membrane is shown in Fig \ref{fig:lifetime}(a) for pump powers varying from 2 mW to 30 mW for the sample also at $\sim$3 K. A clear increase in the initial decay rate from the cavity is observed. We also fit the decay traces to a double exponential with the decay time constants shown in Fig. \ref{fig:lifetime}(b). We observe that both the long and short decay lifetimes vary with pump power, which is consistent with the concept of stimulated emission, as discussed below. We note that the initial decay time constant is 4.0 ms for unpatterned film and 0.73 ms for cavity resonant emission.  The measured total PL emission decay time constant ($\tau$) from the unpatterned film is the parallel sum of the radiative ($\tau_{r}$) and the non-radiative ($\tau_{nr}$) decay time constants. On the other hand for a cavity, the radiative decay time constant is shortened by Purcell factor $F_{p}$, and the measured overall time constant for a cavity is:
\begin{equation}
1/\tau_{cav}=F_{p}/\tau_{r}+1/\tau_{nr}.
\end{equation}
The lower bound of the radiative lifetime of Er in bulk Si$_{3}$N$_{4}$ is 7 ms, as it is the longest total (combined radiative and non-radiative) lifetime observed for such a system \cite{Polman_Er}. Using this figure, we conservatively approximate $F_{p}=6$ at 3 K at the lowest pump power.

\begin{figure}[htbp]
\centering
\includegraphics[width=3in]{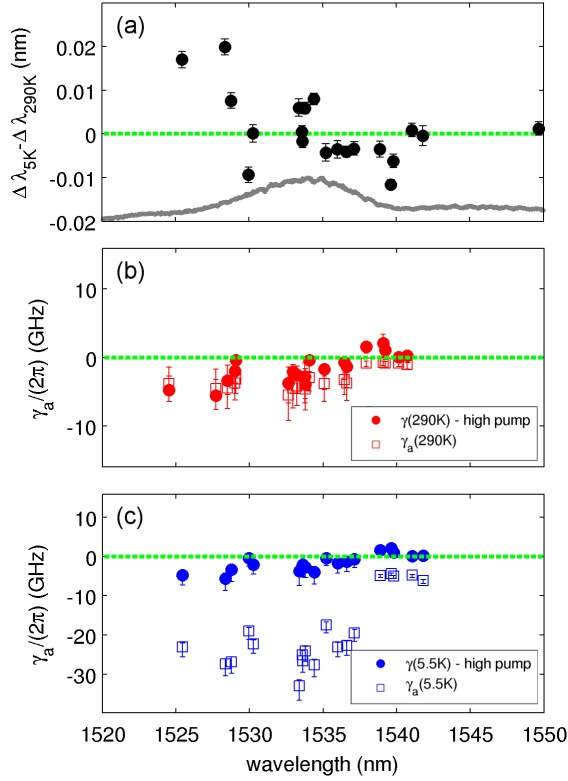}
\caption{(a) The difference in between the cavity linewidth at 5.5 K and 290 K, under high pump power (greater than 40 mW). The scaled and shifted Er spectrum is shown as a reference. (b) The absorption rate achieved at room temperature under high pump power (circles) and low pump power (squares) calculated by use of the cavity $Q$s measured in experiment, with error bounds assuming that the Er homogeneous linewidth at room temperature is between $\eta=4$ and $\eta=8$ times that at 5.5 K. (c) The absorption rate achieved at 5.5 K at high pump power (circles) and low pump power (squares), with the same error bounds as part (b). Regions with positive $\gamma_{a}$ correspond to gain achieved with the system.}
\label{fig:gain}
\end{figure}

We also observe that the $Q$s of the cavities with the type I membrane at 5.5 K and at high pump powers can sometimes exceed the $Q$s of the same cavities at high pump powers at room temperature. We find the difference between the linewidths of the cavity resonances at the two different temperatures at the same high pump power and plot them in Fig. \ref{fig:gain}(a). We observe that a cluster of cavities between 1535 nm and 1539 nm exhibit narrower linewidth at low temperature than at room temperature. Understandably, this range lies on the longer-wavelength side of the Er emission peak, where absorption from Er is lower compared to the shorter-wavelength side of the Er emission peak. The observation of narrower linewidth at 5.5 K indicates that larger gain is achievable at low temperatures than at room temperature. 

In addition, we expect the pump power and temperature dependent absorption rate ($\gamma_{a}(P,T)$) to be related to the observed $Q$-factor of the cavity ($Q_{obs}$) by:
\begin{equation}  
\frac{\omega_{0}}{Q_{obs}} =\frac{\omega_{0}}{Q_{cav}}- \gamma_{a}(P,T),
\label{eqn:qgain}
\end{equation}
where $\omega_{0}$ is the cavity frequency, and $Q_{cav}$ is the intrinsic (no gain or loss) cavity $Q$. Because the light-in light-out (LL) curve does not indicate clear threshold behavior, the system operates well under any lasing conditions, and Eq. \ref{eqn:qgain} is valid. However, as discussed in our previous work \cite{Erbcav_gain}, $\gamma_{a}(T)$ is dependent on the Er homogeneous linewidth, and thus has a strong dependence on temperature. By using the time-resolved data above, we have estimated that the average enhancement of radiative lifetime to be 5 to 7 times stronger at $\sim$3 K than at room temperature, the same as in our previous work \cite{Erbcav_gain}. At room temperature, the homogeneous linewidth of Er in SiN$_{x}$ is estimated to be greater than 1 nm, and is much broader than the cavity linewidth. In that case, the system is in the bad emitter limit \cite{Kimble}, and the spontaneous emission rate enhancement at room temperature is $F(T)\propto Q_{Er}/V_{mode}$, where $Q_{Er}=\omega_{0}/\Delta\omega_{Er}$, and $\omega_{0}$ and $\Delta\omega_{Er}$ are the Er transition frequency and Er homogeneous linewidth, respectively. On the other hand, at cryogenic temperatures, if $\Delta\omega_{Er}$ were to be comparable to or smaller than the cavity linewidth (bad cavity limit), we would expect $F(T)\propto Q_{cav}/V_{mode}$, which in this case would be over 50 times larger than at room temperature. Since we observe a change in Purcell enhancement of only 5 to 7, we confirm that the Er homogeneous linewidth limits Purcell enhancement at cryogenic temperatures as well.

Next, we use the time-resolved data to estimate the Er inversion fraction. We may write for a single cavity that $\gamma_{a}(290$ K$)=\gamma_{a}$, where $\gamma_{a}$ is the cavity dependent absorption rate, and that $\gamma_{a}(5.5$ K$)=\eta\gamma_{a}$, where $\eta$ is the factor by which the Er homogeneous linewidth decreases between room temperature and 5.5 K. By using $\eta=6$ as observed from time-resolved spectroscopy, along with $Q_{obs}$ (deconvolved from the spectrometer response) at low pump powers ($Er^{*}/Er=0$) at 290 K and 5.5 K, we can find $\gamma_{a}$ and $Q_{cav}$ for each cavity using Eq. \ref{eqn:qgain}. We plot $\gamma_{a}$ in Fig. \ref{fig:gain}(b) and (c) for room temperature and 5.5 K, respectively, and the data match well with the expected absorption spectrum of Er. In addition, we find the effective gain (or absorption) rate, namely, $\omega_{0}/Q_{obs}-\omega_{0}/Q_{cav}$, achieved at the cavity resonance wavelength for each cavity. We plot the results for room temperature and 5.5 K in Fig. \ref{fig:gain}(b) and (c), respectively, with the cases of $\eta=4$ and $\eta=8$ as the error bar bounds. As with the data in Fig \ref{fig:gain}(a), we observe that the fraction of inverted Er rises above transparency, i.e., $\gamma_{a}$ equal to or greater than zero (otherwise $\gamma_{a}<0$ denotes absorption loss), for the cavities coupled to the longer wavelength side of the main Er emission peak at both room temperature and 5.5 K. Once again, such an effect matches well with the pump power dependent gain curves of Er in glass \cite{Desurvire_pg232}. The absorption coefficient ($\alpha$) can be calculated from the absorption rate by $\alpha = -\gamma_{a}/(2\pi)/(c\Gamma /n_{eff})$, where for this cavity mode the effective index is $n_{eff}=1.6$ and mode overlap with the active material is $\Gamma = 0.52$. At the Er emission peak, we obtain an absorption rate of $\gamma_{a} = 2\pi \times 6$ GHz, which corresponds to an absorption coefficient of 0.6 cm$^{-1}$ and is consistent with absorption rate of Er doped materials in silicon nanocrystal doped oxide and phosphate glass waveguide systems \cite{Polman_ErP,Han_Erwave}. Similarly, the maximum gain ($\gamma_{a} > 0$) obtained at 5.5 K and 290 K is $\gamma_{a} = 2\pi \times 2$ GHz, which corresponds to $\alpha = -0.22\pm 0.05$ cm$^{-1}$. In general, we observe that at the long wavelength edge of the main Er emission peak, cavities are pumped to transparency.

\begin{figure}[htbp]
\centering
\includegraphics[width=4.5in]{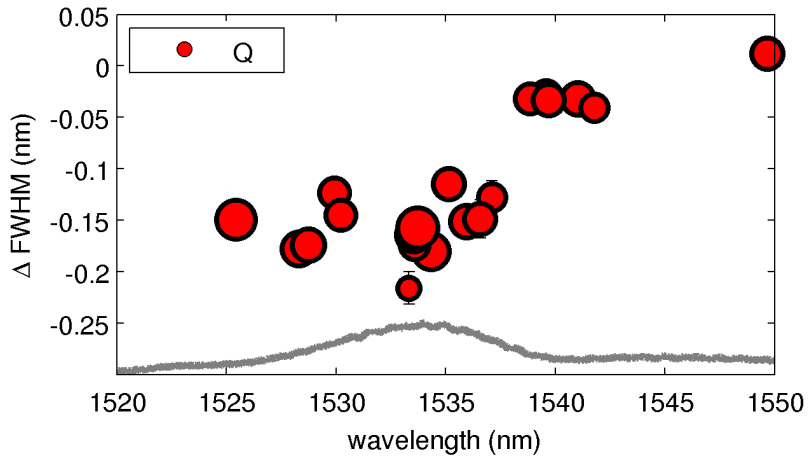}
\caption{(a) The change in linewidth between 5.5 K and room temperature, both measured at low pump powers (below 10 $\mu$W). The size of the points represents the intrinsic cavity $Q$-factor ($Q_{cav}$). The reference marker is for a cavity with $Q_{cav}=12,000$. The scaled and shifted Er spectrum is shown as a reference.}
\label{fig:comp_Qs}
\end{figure}

Finally, we confirm that the Purcell enhancement is degraded by the large homogeneous linewidth of the Er transition at both 5.5 K and room temperature. We plot the change in the cavity linewidth at low pump power between 5.5 K and room temperature for various cavities and simultaneously plot $Q_{cav}$ as the area of the points in Fig. \ref{fig:comp_Qs}. We observe that the change in the linewidth (i.e., the change in absorption) between the two temperatures is not strongly correlated with the intrinsic cavity $Q$-factor, i.e., the size of the points does not increase for larger changes in linewidth. Thus, we observe that the linewidth broadening effect due to Purcell-enhanced absorption at low temperatures is saturated for high-$Q$ cavities. Therefore, minimizing the cavity mode volume while keeping the cavity linewidth comparable to the homogeneous Er linewidth would achieve the maximum Purcell enhancement in nano-cavity structures.

In conclusion, we have demonstrated that high $Q$-factor and low $V_{mode}$ nanobeam cavities enable the observation of stimulated emission and optical transparency in an Er:SiN$_{x}$ material system. We have observed enhanced absorption and gain characteristics in the Er:SiN$_{x}$ compared to our previous work with a hybrid Er:SiN$_{x}$/Si PC cavity \cite{Erbcav_gain}, due to the increased overlap of the cavity mode with the active Er material and the reduction of material losses. Finally, we have studied the pump power dependence of the cavities and observed linewidth narrowing. By comparing the cavity linewidths at room temperature and 5.5 K, and accounting for the enhancement of absorption using time-resolved measurements, we found that cavities have been pumped to transparency. Because cavity heating effects remain significant in such a cavity, material properties of the Er:SiN$_{x}$ layer may need to be adjusted to achieve higher inversion ratios. In addition, cavity designs with higher $Q$ and lower $V_{mode}$ may help achieve lasing by reducing the lasing threshold. In spite of such challenges, the observation of transparency due to stimulated emission in the Er:SiN$_{x}$ system is significant for future designs of lasers and amplifiers based in this material system.

This work was supported in part by the grants from MARCO IFC, AFOSR and the U.S. Air Force MURI program under Award No. FA9550-06-1-0470, on ``Electrically-Pumped Silicon-Based Lasers for Chip-Scale Nanophotonic Systems" supervised by Dr. Gernot Pomrenke. We also thank Sander Dorenbos, Robert Hadfield, and Val Zwiller for providing the superconducting nano-wire single photon detector. PC Devices were fabricated in part at the Stanford Nanofabrication Facility of NNIN supported by the National Science Foundation under Grant ECS-9731293. We also acknowledge support from the Intel (MM) and the NSF Graduate Research (YG) fellowships.

\end{document}